\catcode`\@=11

\newif\if@fewtab\@fewtabtrue

{\count255=\time\divide\count255 by 60
\xdef\hourmin{\number\count255}
\multiply\count255 by-60\advance\count255 by\time
\xdef\hourmin{\hourmin:\ifnum\count255<10 0\fi\the\count255}}
\def\ps@draft{\let\@mkboth\@gobbletwo
    \def\@oddhead{}
    \def\@oddfoot
       {\hbox to 7 cm{$\scriptstyle Draft\ version:\ \draftdate$
       \hfil}\hskip -7cm\hfil\rm\thepage \hfil}
    \def\@evenhead{}\let\@evenfoot\@oddfoot}

\def\ceqno{\global\@fewtabfalse
    \ifcase\@eqcnt \def\@tempa{& & &}\or \def\@tempa{& &}
      \or \def\@tempa{&}
      \or\def\@tempa{}\fi\@tempa
{\rm(\theequation)}}

\def\aeqno#1{\global\@fewtabfalse
    \ifcase\@eqcnt \def\@tempa{& & &}\or \def\@tempa{& &}
      \or \def\@tempa{&}
      \or\def\@tempa{}\fi\@tempa
{\rm(\theequation,#1)}}

\def\label#1{\ifnum\draftcontrol=1
 \global\def\draftnote{$\scriptstyle #1$}\fi
 \@bsphack\if@filesw {\let\thepage\relax
   \def\protect{\noexpand\noexpand\noexpand}%
\xdef\@gtempa{\write\@auxout{\string
      \newlabel{#1}{{\@currentlabel}{\thepage}}}}}\@gtempa
   \if@nobreak \ifvmode\nobreak\fi\fi\fi
  \@esphack}

\def\alabel#1#2{\label{#1}\global\@fewtabfalse
    \ifcase\@eqcnt \def\@tempa{& & &}\or \def\@tempa{& &}
      \or \def\@tempa{&}
      \or\def\@tempa{}\fi\@tempa
{\hbox to 3cm{\phantom{\rm(\theequation,#2)}
\hfil}\hskip -3cm {\rm(\theequation,#2)}}}

\def\clabel#1{\label{#1}\global\@fewtabfalse
    \ifcase\@eqcnt \def\@tempa{& & &}\or \def\@tempa{& &}
      \or \def\@tempa{&}
      \or\def\@tempa{}\fi\@tempa
{\hbox to 3cm{\phantom{\rm(\theequation)}
\hfil}\hskip -3cm{\rm(\theequation)}}}

\def\eqnarray{\def\draftnote{{}}\global\@fewtabtrue
\stepcounter{equation}\let\@currentlabel=\theequation
\global\@eqnswtrue
\global\@eqcnt\z@\tabskip\@centering\let\\=\@eqncr
$$\halign to \displaywidth\bgroup\@eqnsel\hskip\@centering\@eqcnt\z@
  $\displaystyle\tabskip\z@{##}$&\global\@eqcnt\@ne
  \hskip 1\arraycolsep \hfil${##}$\hfil
  &\global\@eqcnt\tw@ \hskip 1\arraycolsep
$\displaystyle\tabskip\z@{##}$
\hfil  \tabskip\@centering&\global\@eqcnt\thr@@\llap{##}\tabskip\z@
\cr}

\def\endeqnarray{\@@eqncr\egroup
      \global\advance\c@equation\m@ne$$\global\@ignoretrue}

\def\@eqnnum{\hbox to 3cm{\phantom{\rm(\theequation)} 
                         \hfil}\hskip -3cm {\rm(\theequation)}}

\def\@@eqncr{\let\@tempa\relax
    \ifcase\@eqcnt \def\@tempa{& & &}\or \def\@tempa{& &}
      \or \def\@tempa{&}
      \or\def\@tempa{}
\fi\@tempa
\if@eqnsw
\if@fewtab\@eqnnum\fi
\stepcounter{equation}\fi\global
\@eqnswtrue\global\@eqcnt\z@\global\@fewtabtrue\cr}

\def\draftcite#1{\ifnum\draftcontrol=1#1\else{}\fi}

\def\@lbibitem[#1]#2{\item{}\hskip -3cm \hbox to 2cm
{\hfil$\scriptstyle\draftcite{#2}$}\hskip
1cm[\@biblabel{#1}]\if@filesw
     {\def\protect##1{\string ##1\space}\immediate
      \write\@auxout{\string\bibcite{#2}{#1}}}\fi\ignorespaces}

\def\@bibitem#1{\item\hskip -3cm \hbox to 2cm
{\hfil $\scriptstyle\draftcite{#1}$}\hskip 1cm
\if@filesw \immediate\write\@auxout
       {\string\bibcite{#1}{\the\value{\@listctr}}}\fi\ignorespaces}

\def\nsection#1{\section{#1}\setcounter{equation}{0}}

\font\tendl=eufm10  scaled \magstep1
\font\sevendl=eufm7 scaled \magstep1
\font\fivedl=eufm5 scaled \magstep1
\font\tengl=eufm10  scaled \magstep1
\font\sevengl=eufm7 scaled \magstep1
\font\fivegl=eufm5 scaled \magstep1

\newfam\dlfam  
\textfont\dlfam=\tendl \scriptfont\dlfam=\sevendl
\scriptscriptfont\dlfam=\fivedl
\newfam\glfam  
\textfont\glfam=\tengl \scriptfont\glfam=\sevengl
\scriptscriptfont\glfam=\fivegl

\def\draftdate{\number\month/\number\day/\number\year\ \ \ \hourmin }

\global\def\draftcontrol{0}
\catcode`\@=12
\def\tilde{\widetilde}
\def\hat{\widehat}
\documentstyle[12pt]{article}

\def\theequation{{\thesection.\arabic{equation}}}

\newcommand{\be}{\begin{eqnarray}}
\newcommand{\en}{\end{eqnarray}\vs 0.5 cm}

\newcommand{\qq}{\begin{eqnarray}}
\newcommand{\qqq}{\end{eqnarray}}

\newcommand{\s}{\hspace{0.05cm}}

\def\ifundefined#1{\expandafter\ifx\csname#1\endcsname\relax}
\makeatletter \ifundefined{new@mathgroup} {} \else   
    \new@mathgroup\sffam
    \new@internalmathalphabet\mathsf\sffam{cmss}{m}{n}
    \def\psf{\fontfamily\sfdefault \fontseries\default@series
        \fontshape\default@shape\selectfont\mathsf}
\fi

\def\citen#1{\if@filesw \immediate\write \@auxout {\string\citation{#1}}\fi%
\@tempcntb\m@ne \let\@h@ld\relax \def\@citea{}%
\@for \@citeb:=#1\do {\@ifundefined {b@\@citeb}%
    {\@h@ld\@citea\@tempcntb\m@ne{\bf ?}%
    \@warning {Citation `\@citeb ' on page \thepage \space undefined}}%
    {\@tempcnta\@tempcntb \advance\@tempcnta\@ne
    \setbox\z@\hbox\bgroup\ifcat0\csname b@\@citeb \endcsname \relax
    \egroup \@tempcntb\number\csname b@\@citeb \endcsname \relax
    \else \egroup \@tempcntb\m@ne \fi \ifnum\@tempcnta=\@tempcntb
    \ifx\@h@ld\relax \edef \@h@ld{\@citea\csname b@\@citeb\endcsname}%
    \else \edef\@h@ld{\hbox{--}\penalty\@highpenalty
    \csname b@\@citeb\endcsname}\fi
    \else \@h@ld\@citea\csname b@\@citeb \endcsname \let\@h@ld\relax \fi}%
\def\@citea{,\penalty\@highpenalty\hskip.13em plus.13em minus.13em}}\@h@ld}
\def\@citex[#1]#2{\@cite{\citen{#2}}{#1}}%
\def\@cite#1#2{\leavevmode\unskip\ifnum\lastpenalty=\z@\penalty\@highpenalty\fi%
  \ [{\multiply\@highpenalty 3 #1%
  \if@tempswa,\penalty\@highpenalty\ #2\fi}]}   %
\makeatother 

\def\be            {\begin{equation}}

\def\cft           {conformal field theory}
\def\cfts          {conformal field theories}
\def\Chi           {{\cal X}}

\def\cym           {Calabi\hy Yau manifold}

\def\cts           {coset theories}

\long\def\del#1    \enddel{}

\def\dyd           {Dynkin diagram}
\def\ee            {\end{equation}}
\def\eE            {{\rm e}}

\def\epop          {extended Poin\-car\'e polynomial}
\newcommand{\erf}[1]{(\ref{#1})}

\def\futnote#1     {\footnote{~#1}\ }

\def\gv            {\mbox{$g_{}^\vee$}}

\def\hil           {\mbox{$ \cal H$}}

\def\hsc           {hermitian symmetric coset}

\def\hy            {$\mbox{-\hspace{-.66 mm}-}$}

\def\ii            {{\rm i}}
\def\insec         {\mbox{${\cal C}_9$}}
\def\insecg        {\mbox{${\cal C}_{3+6n}$}}

\def\jf            {J.\ Fuchs}

\def\jv            {\mbox{$J_{\rm v}$}}

\long\def\labl#1   {\label{#1}\ee\mbox{ }\\[-12 mm]\query{#1}\\[5 mm] }
\def\laful         {{\tilde\lambda}}

\def\lagi          {Landau\hy Ginz\-burg}

\def\lie           {Lie algebra}

\def\ngen          {\mbox{$N_{27}^{}$}}
\def\nagen         {\mbox{$N_{\overline{27}}$}}
\def\nn            {$N=2$\ }

\def\NS            {Neveu\hy Schwarz\ }
\def\onehalf       {\mbox{$\frac12$}}

\def\opop          {ordinary Poin\-car\'e polynomial}

\def\pop           {Poin\-car\'e polynomial}
\def\ptx           {\mbox{${\cal P}(t,x)$}}

\def\qfts          {quantum field theories}
\def\qq            {\mbox{$\xi_\circ$}}

\long\def\query#1{\hskip 0pt{\vadjust{\everypar={}\small\vtop to 0pt{\hbox{}%
     \vskip -13pt\rlap{\hbox to 46.3pc{\hfil{\vtop{\hsize=8pc\tolerance=6000%
     \hfuzz=.5pc\rightskip=0pt plus 3em\noindent#1}}}}\vss}}}}%
\def\R             {Ramond }
\long\def\rank#1   {\mbox{rank}\,#1}

\def\rep           {representation}

\def\rhog          {\mbox{$\rho_g$}}
\def\rhoh          {\mbox{$\rho_h^{}$}}

\def\rgs           {Ramond ground state}

\def\scu           {simple current}
\def\siebq         {\mbox{$\overline{27}$}}

\def\smat          {$S$-matrix}
\def\sod           {\mbox{so(2$d$)}}
\def\sode          {\mbox{so(2$d$)$_1$}}

\def\stot          {\mbox{$S_{{\rm tot}}$}   }

\def\susy          {su\-per\-sym\-me\-try}
\def\Tau           {{\cal T}}
\renewcommand{\theequation}{\thesection.\arabic{equation}}

\def\ue            {\mbox{u(1)}}

\def\wzwt          {WZW theory}
\def\wzwts         {WZW theories}

   \newcommand{\wb}{\,\linebreak[0]} 
   
   \newcommand{\J}[1]     {{{#1}}\vyp}

   \newcommand{\bi}[1]    {\bibitem{#1}}
   
   \newcommand{\Bi}[1]    {\bibitem{#1}}
   
   \newcommand{\Prep}[2]  {preprint {#1}}

   \newcommand{\vyp}[4]   {\ {#1} ({#2}) {#3}} 

   \def\anop  {Ann.\wb Phys.}
   \def\comp  {Com\-mun.\wb Math.\wb Phys.}

   \def\ijmp  {Int.\wb J.\wb Mod.\wb Phys.\ A}

   \def\mpla  {Mod.\wb Phys.\wb Lett.\ A}

   \def\nupb  {Nucl.\wb Phys.\ B}
   
   \def\phlb  {Phys.\wb Lett.\ B}

   \def\rmap  {Rev.\wb Math.\wb Phys.}

\renewcommand{\labl}[1]{\label{#1} \ee}
\begin{document}
\
\begin{flushright}  
{\bf hep-th/9311168} \\
{\bf HD-THEP-93-43}
\end{flushright}
\begin{center}
\vskip 0.9cm

{\Large{\bf{POINCARE POLYNOMIALS AND LEVEL RANK}}}
\vskip 0.3cm

{\Large{\bf DUALITIES IN THE $N=2$ COSET}}
\vskip 0.3cm

{\Large{\bf CONSTRUCTION} 
\footnote{\s Invited talk given at the III. International
Conference on Mathematical Physics, String Theory
and Quantum Gravity, Alushta, Ukraine, June 1993. To appear in Theor.\ Math.\
Phys.}}
\vskip 0.5cm

\vskip 0.7cm

{\large{Christoph Schweigert}}
\footnote{\s supported by Studienstiftung des deutschen Volkes}
\vskip 0.2cm

Institut f\"ur theoretische Physik,

Philosophenweg 16

69120 Heidelberg, Germany

\vskip 0.5cm

\end{center}
\date{ }

\vskip 1.6cm

\begin{abstract}
\noindent
We review the coset construction of conformal field theories; the emphasis
is on the construction of the Hilbert spaces for these models, especially if
fixed points occur. This is applied to the $N=2$ superconformal cosets
constructed by Kazama and Suzuki. To calculate heterotic string spectra we
reformulate the Gepner construction in terms of simple currents and introduce
the so-called extended Poincar\'e polynomial.
We finally comment on the various equivalences arising between models of this
class, which can be expressed as level rank dualities.

\end{abstract}
\vskip 1.3cm

\nsection{Introduction}
Conformal field theory in general and $N=2$ superconformal field theories
in particular have provided a deeper understanding of many issues
in both mathematics and physics.

In physics, string theory is still one of the most exiting paths towards a
quantization of gravity and also implications in the realm of statistical
physics, e.g.\ for universality classes of two-dimensional critical behaviour,
have been established. Moreover, deep and beautiful relations to other
types of quantum field theories, e.g.\ Toda theories, Chern-Simons
theories and integrable systems have been unraveled. Conformal field theories
provide also an important testing ground for the algebraic theory of
superselection sectors.

Mathematics has largely benefitted  from \cft\ as well: it gave rise
to a new and rich interplay between various branches: algebraic and
differential geometry, number theory, infinite dimensional \lie s,
the theory of $C^*$ algebras, commutative algebra, just to enumerate some
of them.

The interest in \nn\ superconformal theories was initially motivated in string
theory
by the fact that -- together with charge quantization -- $N=2$ superconformal
symmetry on the world sheet yields a space time supersymmetric string spectrum,
but nowadays an independent motivation to study these models comes from their
beautiful intrinsic structure and their deep connection to other objects
in mathematical physics, e.g.\  to two-dimensional topological \qfts\ and
possibly even to general \cfts.

Among the various ways to construct these models -- non-linear sigma models
with the target space a \cym , (infra-red) fixed points of the
renormalization group flow on \lagi\ potentials etc.\ -- exactly solvable
models
are distinguished by two important properties.
Not only can one calculate in these models -- at least in principle --
exactly, i.e.\ non-perturbatively correlation functions, but,
what is even more fundamental, the full field content of these models is
known. This allows for an explicit calculation of the
behaviour under modular transformations, giving complete information on
quantum dimensions and fusion rules.
So it is in these models that one hopes to really identify the fundamental
symmetries of quantum physics.
Our main interest in this talk will be in \cts , where the mathematical
framework is the theory of affine \lie s, but, of course, there
are also other completely solvable formulations, e.g.\ the Coulomb gas
approach.

\nsection{The coset construction}

Starting from a complex, affine (untwisted) Kac-Moody algebra at level $k$ :
\be
[J^a_n, J^b_m] = f^{ab}_c J^c_{n+m} + k m \, \delta_{m+n, 0} \kappa^{ab}
\labl{}
we obtain by the Sugawara construction
\be L_n := \frac{1}{2k + 2 \gv} \sum_{m, a, b} \kappa_{a b}
: J^a_{m+n} J^b_{-m} : \ee
the Virasoro algebra:
\be [L_n , L_m ] = (n-m) L_{n+m} + \frac{c}{12} (n^3 -n) \delta_{n+m, 0}
\labl{vir}
with central charge $ c = \frac{k \dim g}{ k + g_{}^\vee} $.
Many quantities of interest of a \wzwt\ can be entirely described in terms
of the simple Lie algebra $g$ that is generated by the zero mode currents.
So in the equations above $\kappa$ denotes the Killing form of $g$ and \gv\ the
dual Coxeter number; colons indicate normal ordering. Imposing unitarity,
which is a natural requirement in string theory, but not necessarily in
statistical mechanics, we end up with a rational \wzwt\ at integer level $k$.

The Virasoro central charges obtained this way always obey
$c \geq \mbox{rank}\, g $, with equality exactly for simply-laced \lie s at
level $k=1$; so the various minimal series cannot be obtained
this way.  This was one motivation for the coset construction
\cite{goko}, which turned out to be a powerful tool to construct new \cfts.

Consider a subalgebra $h$ embedded in $g$. It is easy to show that
\be L^{g/h}_n := L^g_n - L^h_n \ee
generates a Virasoro algebra with central charge $c^{g/h} = c^g - c^h$  .

While on the level of the symmetry algebra the construction is
straightforward, we have to emphasize the fact that we are still far away from
having constructed a conformal field theory. The question we have to
worry about is: what is the Hilbert space of the theory? As it turns out,
this is a highly non-trivial question and at present no completely satisfactory
answer to it is known.

A first guess is to decompose any irreducible unitary representation
$\hil^g_\Lambda$ of $g$ in irreducible representations $\hil^h_\lambda$ of the
subalgebra $h$:
\be
\hil^g_\Lambda = \bigoplus_\lambda \hil^\Lambda_\lambda \otimes \hil^h_\lambda
,
\ee
what leads to the following relation between the characters:
\be \chi^g_\Lambda (\tau )= \sum_\lambda b^\Lambda_\lambda(\tau )
 \chi^h_\lambda (\tau )  . \ee
The Hilbert space of the coset is then supposed to be all
$\hil^\Lambda_\lambda$ , its characters are the {\em branching functions}
$b^{\Lambda}_\lambda$ .

But right here we start running into problems. To see what happens let us
have a look at the critical Ising model, i.e.\ the minimal model with
$c=\onehalf$, which has three primary fields: the vacuum
with conformal weight $\Delta =0$, the twist field $\sigma$ with $\Delta
= 1/16$ and the energy operator $\epsilon$ with $\Delta = \onehalf$.
It can be described by the coset:
\be \frac{ (A_1)_1 \oplus (A_1)_1 }{ (A_1)_2}  . \labl{isi}
Here subscripts are used to indicate the levels of the various algebras.
A field is labeled by three quantum numbers: $\Phi^{p q}_r$ , where unitarity
restricts $p$ and $q$ to be $0$ or $1$, and $0 \leq r \leq 2$.

We immediately see that group theoretical selection rules for the coupling
of the two $A_1$ representations force all fields with $p+q-r \neq 0 \bmod 2$
to vanish. A second glance at the branching functions of the coset \erf{isi}
reveals that, apparently, each field occurs {\em twice} in the spectrum.
The vacuum, e.g.\ is represented by $\Phi^{0 0}_0$ or $\Phi^{1 1}_2$,
$\sigma$ by $\Phi^{0 1}_1$ or $\Phi^{10}_1$, and $\epsilon$ by $\Phi^{1 1}_0$
or $\Phi^{0 0}_2$. A more careful analysis reveals that moreover modular
invariance is spoiled.

Here a new requirement we impose on a conformal field theory enters: modular
invariance.
In string theory, in a string loop expansion \`a la Polyakov,
modular invariance is used to factor out the `large' diffeomorphisms
in the diffeomorphism group. In fact, we require even more than just modular
invariance of the partition function; we also require that the characters form
an unitary representation of the modular group.

The situation encountered in the coset representation of the critical Ising
model generalizes: some branching functions vanish while
other non-vanishing branching functions turn out to be identical. There are
three possible reasons for a branching function to vanish: group theoretical
selection rules like in the case of the Ising model, the occurrence of
null states \footnote{This is the case for conformal embeddings (which yield
trivial cosets with $c=0$) and also for the `maverick' cosets described in
\cite{dujo2}.}
and a slightly more complicated combination argument \cite{scya6}, but for
which
no example is known. On the other hand, in general, \smat\ elements
between vanishing and non-vanishing branching functions do not vanish, so
it is impossible to simply delete the corresponding rows and lines in the
\smat\ without spoiling unitarity of the \smat.

Before explaining the way out of this situation, it is appropriate to recall
some results from the theory of \scu s (for a review see \cite{scya6}).
A \scu\ $J$ is a primary field in the theory for which the fusion product
with any field of the theory yields just one other field:
\be J \star \Phi =  \Phi '  .\label{smpl}  \ee
We define the monodromy charge $Q$ of $\Phi$ relative to $J$ as $Q= \Delta(J) +
\Delta(\Phi) - \Delta(\Phi')$.
In the case of \wzwts\ $Q$ can be shown to be simply the conjugacy class.

Simple currents give rise to modular invariants which can be different from
the diagonal one. Important for our purposes is the case of
integer spin \scu s. In this case a non-trivial modular invariant exists
in which only fields with vanishing monodromy charge occur. It is given
by:
\be
Z = \sum_{a, Q(a) = 0 } \frac{N_0}{N_a} \quad \left|\sum_{i = 0}^{N_a-1}
\chi_{J^i a}\right|^2
\labl{sinv}
Here $N_a$ is the length of the orbit of the simple current $J$, $N_0$ is
the length on the orbit which contains the vacuum, i.e.\ $N_0$ is the order of
\
the simple current.

These invariants are suited to take care of group theoretical selection rules:
non-vanishing fields can be characterized by the fact that they have
vanishing monodromy charge relative to a subgroup of the group of simple
currents of $g$ tensored with (the complement of) $h$. \footnote{The modular
invariants of the maverick cosets are in general {\em not} simple current
invariants.}

We find thus that the true fields of the theory are the {\em orbits}
of the identification group; in the literature this is referred to as `field
identification' \cite{sche3,gepn8}. In fact, this seems to be a generic feature
of reduction
procedures; e.g.\ in a system with first class constraints it is well
known that we have to factor out the action of a `gauge' group, too, to
obtain a consistent physical system.

In our case problems arise if orbits of different lengths occur. Namely, we
want to keep just one representative of of every orbit to have a unique
vacuum; in other words, we have to divide $Z$ by $N_0^2$. But this
would lead to non-integer coefficients in the partition function of shorter
orbits, what is incompatible with the interpretation of $Z$ as a partition
function. The shorter orbits which are termed `fixed points' require thus a
special treatment, and here we clearly need some additional input.

The idea suggested by the prefactors of the complete squares in \erf{sinv}\
is that every fixed point of length $N_f < N_0$ has to be resolved in
$N_0 / N_f$ distinct physical fields. This procedure possibly introduces
some arbitrariness, as \smat\ and characters for the individual physical fields
are a priori unknown.

For the \smat\ elements involving fixed points we make the following
ansatz \cite{scya6}:
\be
\tilde{S}_{f_i g_j} = \frac{N_f N_g}{N_0} S_{f g} + \Gamma^{f g}_{ij}, \ee
where the indices $i,j$ count resolved fields and $S$ is the naive \smat.
Modular invariance can be shown to imply the following sum rules for the
characters $\Chi_{f_i}$ and the \smat\ elements
of fixed points:
\begin{eqnarray}
\sum_i \Chi_{f_i}  &=&  \Chi_f    \label{suru} \\
\sum_i \Gamma^{fg}_{ij}  = & 0 & =  \sum_j \Gamma^{f g}_{i j}
\end{eqnarray}
In most cases the $\Gamma$ matrices and also the character modifications
needed to fulfill \erf{suru}\ can be described in terms of
a different \wzwt , which is usually called `fixed point theory'. A list
of these fixed point theories can be found in \cite{scya6}.

We emphasize that only after having found a consistent resolution of the
fixed points we have really constructed a conformal field theory.
Unfortunately, no general results concerning existence or uniqueness of the
resolution are known.

\nsection{\nn\ \cts}

In the sequel we will focus our attention on a special subclass of cosets
models, namely the \nn\ coset models constructed by Kazama and Suzuki
\cite{kasu,kasu2}. As is well known, the  \nn\ algebra is generated by the
stress energy tensor $T$, two spin $3/2$ supercurrents $G^{\pm}$ and one
spin 1 \ue-current $J$. With respect to $J$ the supercurrents $G^{\pm}$ have
charge $\pm 1$ . For the following considerations it is convenient to define
\be
G_{}^{(1)}  :=   \frac 1 {\sqrt 2} (G^+ + G^- )   \qquad
G_{}^{(2)}  :=   \frac 1 {\ii \sqrt 2} (G^+ - G^- )   .
\ee
Then $T$ and $G_{}^{(1)}$ or $G_{}^{(2)}$ generate an $N=1$ superconformal
algebra.

We are now going to describe the construction proposed by Kazama and Suzuki
\cite{kasu2}.
Given a reductive  subalgebra $h$ of a simple \lie\ $g$, the
embedding $h \hookrightarrow g$ induces also an embedding (`tangent
space embedding') of $h \hookrightarrow \sod $ , where $2d = \dim g -
\dim h$ . We can thus consider the diagonal embedding
\be h \hookrightarrow g_k \oplus  \sode ,  \labl{kasuform}
where subscripts again denote levels.
The choice of this embedding can be motivated by a supersymmetric extension of
the coset construction using super Kac-Moody algebras. In particular, the \sod\
part corresponds to bosonized free fermions, and it is clear that the
corresponding modular invariant should be chosen.

These models can be shown to have always $N=1$ \susy, where the supercurrent
is given by
\be G_{}^{(1)} = G^{g/h} = G^g - G^h = \frac2k (\kappa_{\bar a \bar b}
:j^{\bar a} \hat J^{\bar b}: - \frac{\ii}{3k} f_{\bar a \bar b \bar c}
: j^{\bar a} j^{\bar b} j^{\bar c} :). \ee \label{sucu}
Here $\kappa$ is the Killing form, $f$ are the
structure constants of $g$. The bar over the indices indicates that the sum is
only over elements in the orthogonal complement of $h$ in $g$ relative to the
Killing form. $\hat J$ are the purely bosonic currents and $j^{\bar a}$ the
fermionic currents transforming in the vector representation of \sod.

One may now ask in which cases the symmetry algebra can be enlarged to an \nn
algebra.
To investigate this question we make the most general ansatz in terms of
normal ordered products of fields for a spin $3/2$ current and plug it
into the \nn\ algebra:
\be G_{}^{(2)}  = \frac2k (h_{\bar a \bar b} : j^{\bar a}
\hat J^{\bar b}: - \frac{\ii}{3k} S_{\bar a \bar b \bar c} : j^{\bar a}
 j^{\bar b} j^{\bar c} :). \ee \label{sucu2}

This leads to a system of algebraic equations for
$h$ and $S$. This system was analyzed in \cite{kasu} in the case of regular
embeddings $h \hookrightarrow g$, and for special embeddings in \cite{schW}.
The resulting classification can be summarized as follows: a subalgebra $h$
of $g$ yields an \nn\ supersymmetric coset of the form \erf{kasuform} if and
only if the Dynkin diagram of $h$ can be obtained from the (non-extended)
Dynkin diagram of $g$ by removing at least one node. The situation turns out
to be particularly simple if the node corresponds to a cominimal weight,
i.e.\ a weight with Coxeter label equal to one. One can
easily show \cite{fuSc} that then the corresponding homogeneous space is
hermitian symmetric. In fact most of the examples considered in the literature
belong to this special subclass, in particular to the so-called projective
cosets
\be \frac{ \mbox{SU(n+1)}_k \oplus \mbox{SO(2n)}_1}{\mbox{SU(n)}_{k+1}
\oplus \mbox{U(1)} } \quad . \ee
The \nn\ minimal models can be recovered from the projective cosets
by setting $n=1$.

At this point one may ask whether all rational \nn\ \cfts\ can be represented
in the form \erf{kasuform}. We cannot answer this question. Actually
one representation of an \nn\ coset is known that is not of this form, namely
at $c=1$:
\be
\frac{(A_1)_2 \oplus (A_1)_2}{(A_1)_4^D} \labl{ausn}
But, as this model is a minimal model, it can alternatively also be described
in the form \erf{kasuform}.

In \cite{fuSc} all tensor products of \nn\ cosets that have central charge
$c=9$ and are thus suitable for the inner sector of a heterotic string
compactification have been classified.
To state the result of our classification we count cosets; as different modular
invariants of the same coset yield different conformal field theories (and thus
superstring vacua) the number of the latter is certainly higher.
Unfortunately, it is unknown, as a classification of modular invariants of
general \wzwts\ is still missing.

There are 168 tensor products of minimal models, 190 tensor products of
projective cosets and minimal models, 123 tensor products involving other
\hsc s \cite{sche3} and 198 tensor products involving at least one coset
that is not a \hsc\ \cite{fuSc}. Here some trivial group theoretical
identifications have already been taken into account, such as $C_2 \equiv B_2
$.
Using non-trivial relations, such as the level rank dualities discussed below
(compare section \ref{slevr} and \cite{fuSc2}) the number of distinct coset
\cfts\ is reduced even further, e.g. for those involving non-\hsc s from 198
to 112.

\nsection{Explicit Calculations in \nn \cts}

As should be clear from the preceding discussion, an \nn\ coset has to be
given a sense as a \cft , just like any other coset theory. Therefore first
the \ue\ part has to be specified and the levels of the simple ideals $h_i$
of $h$. The latter can be shown to be $ k_i = I_i (k + \gv ) - h_i^{\rm v}$,
where $I_i$ is the index of the embedding $h_i \hookrightarrow g$.

Then group theoretical selection rules or, equivalently, the identification
rules have to be determined.
In \cite{fuSc} a simple formula for the number of identification currents
was derived which is useful to check the completeness of a set of selection
rules:
\be | {\cal G}_{id} | = Q_{i_0} \prod_i I_C (\hat h_i ) ,  \ee
where $Q_{i_0}$ is the \ue\ charge of the missing simple root (whose node
in the \dyd\ has been deleted) and $I_C$ is the {\em index of connection},
which is equal to the number of conjugacy classes.

Interesting quantities are the number \ngen\ of massless generations
and \nagen\ of anti-generations of the heterotic string compactification
which has a $c=9$ tensor product of coset theories in its inner sector.
For simplicity, we restrict ourselves to the diagonal modular invariant.

Several problems have to be overcome:
firstly, in coset models, only the fractional part of the conformal
weight can be easily obtained. The integer part can be obtained in principle
from a character decomposition using the Weyl-Kac character formula, but
this is extremely tedious and in practice hardly feasible. A way out is to
work with \rgs s which, due to spectral flow, provide equivalent information.
An index-like argument \cite{levw} shows that a complete set of
representatives $\Phi^{\Lambda x}_\laful$ for all \rgs s is given by the
formula
\be \laful = w(\Lambda + \rhog ) - \rhoh  ,  \ee \label{lvwfor}
where $w$ runs over all elements of the Weyl group of $g$ such that $\laful$
is a highest weight of $h$, including an \ue-weight. As this formula has been
derived from the Weyl-Kac character formula, it automatically takes care of
null states. For arbitrary states null states are a severe problem; this makes
e.g.\ the determination of $E_6$ singlets very cumbersome in practice. In fact,
their number has only been determined for tensor products of minimal models,
where the representation theory of the \nn\ algebra provides an independent
powerful handle on null states.

Also, in general, the superconformal charge $q$ can be read off easily only
$\bmod \, 2$; only for \rgs s the following formula \cite{ekmy,gepn11,fuSc}
holds:
\be q = \frac d 2 - l(w) - \frac{\xi_0 Q}{k + \gv} ,  \labl{laefo}
where $l(w)$ is the length of the Weyl group element given by \erf{lvwfor} and
$\xi_0$ some factor of proportionality.
Using equation \erf{laefo} it was shown \cite{fuSc} that for any known \nn\
coset theory the set of \rgs s is symmetric under charge conjugation.

Once we know all \rgs s or, equivalently, due to chiral flow, the ring of
chiral primary fields, we can in principle implement the method of $\beta$
vectors \cite{gepn3} . In practice, this turns out to be rather
inconvenient, but there is a different approach, which has the additional
benefit to provide also the very important insight that the spectra do
{\em not} depend on the details of the resolution procedure for the fixed
points, contrary to what one might expect.

In \cite{butu} modular invariance was used to express
the Euler number $\chi = 2 ( \ngen - \nagen ) $ in terms of the \pop\ $P$:
\be \chi = \frac{1}{M} \sum_{r,s = 0}^{M-1} P(\eE ^{2 \pi {\rm d}(r,s) / M} .)
\labl{eul}
Here $M$ is the smallest common denominator of all \ue-charges in the chiral
ring and ${\rm d}(\cdot,\cdot)$ stands for the largest common divisor of
two integers.

Fixed points of the identification group have to be carefully taken into
account, of course. However, one can show that any orbit containing at least
one representative of  \erf{lvwfor} yields after resolution exactly one
\rgs , no matter how long the orbit is. It is important to realize that this
result holds irrespective of the details of the resolution procedure and that,
as a consequence of \erf{eul}, $\chi$ does not depend on these details either.

\nsection{The \epop }

The method of the \epop\ finally allows for a separate calculation of
\ngen\ and \nagen .

If we want to build out of a tensor product \insec\ of  \nn super\cfts\ with
$c=9$ a heterotic string theory, we have to perform
several projections. We will sketch below how this can be described in terms
of simple currents and explain the resulting prescription encoded in the
`\epop'. In a second step we shall comment on the case $c = 3 + 6n$ . This
case is of much practical interest as we have to resort to it
in some cases to remove ambiguities in the resolution of fixed points.

After splitting off the contribution of the bosonic space time coordinates
and applying the bosonic string map \cite{enns}, we can describe the heterotic
string in a \cft\ language as the tensor product
\be (D_5)_1 \oplus (E_8)_1 \oplus \insec\     . \labl{cft}
The first two factors will provide for the right movers the gauge multiplet,
for the left movers, they describe the contribution of the
fermionic coordinate fields. As the only purpose of the $E_8$ factor is
to provide a phase in the \smat\ such that the fermions are
correctly reproduced, we will drop it in our discussion from now on.

First, to obtain supersymmetry on the world sheet, we have to align the
boundary conditions in the various theories such that all fields are either
\R\ or \NS. Therefore, it is important to
note that $T_F := \Phi^{0 v}_0$ is a spin $1/2$ simple current of order 2 that
is present in any \nn\ coset theory. Its monodromy charge is $0$ for fields in
the \NS sector and $1/2$ in the \R sector.

Alignment is thus equivalent to enlarging the chiral algebra by all bilinears
$T_F (i) T_F (j)$ (which have conformal dimension 3). In the $D_5$ part we set
$T_F(0) := v$, which is, just like any other field of $(D_5)_1$, a simple
current.

Space time supersymmetry requires the projection on even \footnote{Here we
formulate the condition {\em after} applying the bosonic string map, what
explains the difference to what the reader might expect, namely projection on
odd values \cite{gepn3}.} values of the
\ue\ charges \cite{gepn3}. To implement this projection we note
that the \rgs\ $R_0$ with highest \ue\ charge is always a \scu\ with conformal
dimension $h = \frac{c}{24}$. One can show \cite{fuSc} that there is always
one representative of $R_0$ of the form $\Phi^{0 s}_{0 Q_s}$; from this
explicit form its monodromy charge can be easily seen to be half of the
superconformal charge. The desired projection it is thus equivalent to
including the integer spin simple current $S_{\rm tot} := (s, R_0)$ in the
chiral algebra. Here $s$ is
the spinor simple current of $(D_5)_1$.  $S_{\rm tot}$ has been termed spinor
current in \cite{sche3}. We will see below that its presence in the chiral
algebra assures the existence of a space time gravitino in the corresponding
heterotic string spectrum.

In a \cft\ language a heterotic string theory thus amounts to a \cft\ \erf{cft}
with the modular invariant generated by the integer spin simple current
\stot\ and all bilinear combinations $T_F(i) T_F(j)$.

It is now easy to recover the massless spectrum of the heterotic string.  To
obtain the proper interpretation we recall that in one chiral sector of the
theory, e.g.\ for left movers, we have to apply the bosonic string map: the
$D_5 \oplus E_8$ part is mapped on a so($2)_1$ theory by interchanging vector
and scalar and changing the sign of the spinor and conjugate spinor
representation in the partition function. This map preserves the modular
transformation properties and allows for a description of the fermionic
coordinates of the string.

As we work in a purely bosonic description, fields are massless if $\Delta
= \bar \Delta = 1$. Let us first explain how in this formulation the generic
part of the string spectrum arises which provides the supergauge-
and supergravity-multiplets. Two fields that occur in any \nn\ theory in the
inner sector are the vacuum and the two \rgs s with highest and lowest
\ue-charge. The massless right moving fields that are tensored with the vacuum
of the inner sector have $\Delta =1$ and, due to the charge selection rule,
$q = 0, \pm 2$. These conditions are
fulfilled for the currents of $E_8 \oplus D_5$ and the transverse bosons. In
the modular invariant described above these fields are paired with the
following left movers: $(v,0)$ what yields for the transverse bosons the
graviton (as well as an antisymmetric tensor and the dilaton as the trace) and
for the currents the gauge multiplets. Applying \stot\ in the left moving
sector yields the superpartners of the gauge bosons and the graviton.

In the right moving sector, we also find in the complete square of the
identity the fields \stot\ and $\stot^{\dagger}$, as well as the $0$ of $D_5$
tensored with the \ue-current of the \nn\ algebra. According to the well known
branching of the adjoint representation of $E_6$ to the adjoint
representation, the spinor, conjugate spinor and scalar of $D_5$, these fields
extend the gauge symmetry from $E_8 \oplus D_5$ to $E_8 \oplus E_6$. In
particular cases, if more fields are present, one can even further extend both
the gauge symmetry for right movers and the supersymmetry for the left movers.

To explain how massless (anti-)generations transforming in the $27$ resp.\
\siebq\ representations of $E_6$ arise, we remark that massless states that
are vectors of $D_5$ have $\Delta = \onehalf$ and $q=\pm 1$ in \insec, i.e.
they are (anti-)chiral fields.
Acting twice with $\stot^{\dagger}$ on the vector tensored with a chiral
primary field with $q=1$ yields a spinor tensored with a \rgs\ and in a
second step $0$ tensored with an anti-chiral state with $q=-2$; these
states combine in a $27$ of $E_6$. Starting with an anti-chiral field
and applying \stot\ instead we obtain  states transforming in a \siebq\
of $E_6$. These states can be paired with spinors or conjugate spinors in the
left moving sector; together they give rise to the generations and
anti-generations and their $CPT$ conjugates.

To extract information on the spectra we introduce the following notation:
denote by $h^{p,q}$ the number of fields
which are in both the left and the right moving part of \insec\ chiral
primaries and have superconformal charge $p$ resp.\ $q$; $p,q$ are integers
smaller than $d:= c/3$. These numbers can be seen as analogues to the
Hodge numbers of a Calabi-Yau threefold. In fact, we find the usual symmetries:
$h^{p,q}=h^{q,p}$, as we started from a left right symmetric invariant, and
$h^{p,q}=h^{d-q,d-p}$, due to the conjugation symmetry on the chiral ring.
Note that if the vacuum is not paired with any chiral primary field other
than the unique
chiral primary field with $q = \frac c 3$, we have $h^{0,1}=h^{0,2} = 0$;
in the corresponding heterotic string compactification neither gauge
symmetry nor space time supersymmetry is extended. As this is the most
interesting case we will restrict ourselves to it from now on.
The Euler number is given as usual $\chi := \sum (-1)^{p+q} h^{p,q}$.

The discussion above shows that the number \ngen\ of massless generations
transforming in the $27$ representation of $E_6$ is equal
\footnote{Our notation is different from the one used for Calabi-Yau
manifolds: there the superconformal charge in both sectors is defined with a
relative minus sign, so the number of generations corresponds to the Hodge
number $h^{1,-1}=h^{1,2}$ of the manifold.} to $h^{1,1}$, or equivalently
to the number of fields in the theory, which
are in both sectors spinors of $D_5$ tensored with a \rgs\ with superconformal
charge $-\onehalf$. The massless anti-generations \nagen\ transforming in
the $\siebq$ of $E_6$ can be correspondingly characterized by the fields
which are spinors and \rgs\ with charge $-\onehalf$ in one sector and
conjugate spinors and \rgs\ with charge $+\onehalf$ in the other sector.

We are thus interested in the structure of the relevant simple current orbits.
Let us first look at the orbits of \stot: as we are only interested in the
massless spectrum we start with an arbitrary \rgs\ $(s, R^{(1)}, \ldots$) .
Suppose now that, on the orbit, we encounter $( \jv^{\epsilon_0} {\rm v} ,
T_F^{\epsilon_i} R_i ') $, where
$\epsilon_i$ is 0 or 1 . This state -- which is massive unless all $\epsilon_i$
vanish -- is paired in the \scu\ invariant with the original state in the
other sector of the theory.  But the chiral algebra contains also all
bilinears of the form $(\jv , T_F(i))$: we thus find within the same complete
square of the partition function the corresponding massless state, for
which all $\epsilon_i$ vanish, too. If the $D_5$ part is a spinor this yields a
generation; conjugate spinors correspond to anti-generations.

The information on the orbit of \stot\ is very conveniently encoded in the
\epop \cite{sche3}. To start with, we define it on each factor of the tensor
product separately. As any simple current has finite order, the orbit has some
periodicity which
we first factor out for convenience: for any \rgs\ $R$ we define $N_R$ to be
the
smallest power of the spinor current such that $(S^{N_R}) R$ is equal to $R$ or
$T_F R$. We define $\epsilon(R)$ to be $+1$ in the first and $-1$ in the
second case. The \epop\ is now defined as :
\be \ptx = \sum_R \frac{t^q}{1 - \epsilon (R) x^{N_R} } [
\sum_{m \in \cal F_+} x^m -  \sum_{n \in \cal F_-} x^n ]. \ee
The sum is over all \rgs s $R$, $q$ is the superconformal charge of the chiral
primary field connected via spectral flow. The sets ${\cal
F}_\pm$ are defined by the prescription: $m \in {\cal F_+} $ iff
$(\stot)^m R $ is a \rgs\ and $n \in {\cal F_-} $ iff $(\stot)^n R $ is $T_F$
applied to a  \rgs; in particular all $m, n$ are even.
The \epop\ is {\em not} a polynomial in the new variable $x$, but rather a
series with periodic coefficients. We remark that we recover the \opop\ as
${\cal P}(t,0)$.

We obtain the \epop\ for a tensor product by the following multiplication:
given the \epop s ${\cal P}_i(t,x_i)$ of the factors, first perform the
ordinary product of polynomials and then delete all terms in which the powers
of the $x_i$ do not coincide. This procedure implements the simple observation
that, in order to have a \rgs\ of the tensor product, we need \rgs s in each
factor of the theory.

The statements about the corresponding string compactification can
be rephrased in terms of the \epop. First, note that our assumption
that the symmetry is not enlarged translates into the requirement that the
polynomial in $x$ multiplying $t^0$ is equal to $1+ x^2$. \ngen\ and \nagen\
can be read off from the polynomial in $x$ multiplying $t^1$: let
$p(x) = \sum a_m x^m $ denote one period of this series. As the action of any
of the bilinears $(v,T_F(i) )$ and of $(\stot)^2$ changes the conjugacy class
in
the $D_5$ theory we find generations if $a_m > 0$ and $m = 0 \bmod 4$ or
$a_m < 0$ and $m = 2 \bmod 4$; the other cases correspond to anti-generations.
Put differently we find $\ngen + \nagen = \sum |a_m|$ and $\ngen - \nagen =
p(\ii)$.

The formalism of the \epop\ allows for an easy calculation \cite{sche3,fuSc}
of \ngen\ and \nagen\ even if fixed points are present. Namely, after
writing down those parts of the \epop\ which are not affected by
fixed points and taking into account some evident structure of the
fixed points we are left with only a few candidates for the \epop s. As
pointed out above, the \opop\ and hence the Euler number do not depend on the
resolution procedure. For all possible \epop s of a given theory we calculate
$\chi$ from \ngen\ and \nagen . If we do not get the correct Euler number
\erf{eul}, we can exclude the candidate. This works surprisingly well. But, as
the Hodge numbers turn out to be relatively robust against changes of the
parameters left in the \epop, it is important to have many tensor products
in which a given model appears in order to have enough consistency conditions
to single out the true \epop .

This is our main motivation to generalize this formalism to tensor
products of cosets with conformal charge $c = 3 + 6n$. Here in general, we have
no string interpretation at hand, so we can replace the $D_5$ factor by some
other $D_d$
factor. However, we have to require that the current $(s,R_0)$, with
conformal weight $d/8 + c/24$ has integer spin. This fixes $d$ to
$d= -2n -1 \bmod 8$. (We recover the previous situation for $d=5, n=1$.) It
is important to note that, as the $S$ matrices of $D_d$ and $D_{d+4}$
coincide, the choice of $d$ does not affect the fusion rules. (Note however,
that the $T$ matrices coincide only for $D_d$ and $D_{d+24}$.)

We now implement analogous projections, i.e.\ take the simple current invariant
induced by all bilinears in the $T_F(i)$ and $(s,R_0)$, and obtain
the \epop\ by exactly the same prescription as in the $c=9$ case. Again
massless states that are spinors or conjugate spinors in $D_d$ are  \rgs s of
\insecg. The charge selection rule implies that states paired with spinors
have superconformal charge $q \equiv -\onehalf \bmod 2$ and for conjugate
spinors $+\onehalf \bmod 2$.
The chiral primary fields connected via spectral flow have thus charge
$q \equiv n \bmod 2$ for spinors resp.\ $n+1$ for conjugate spinors.
This shows that we can recover the Euler number from
the polynomials multiplying all odd powers of $t$ in the \epop\ and summing
up all contributions. Comparing this result with the result of \erf{eul},
which was derived in \cite{butu} for all $c=3+6n$, we obtain
new consistency conditions on the coefficients in the \epop\ that
can arise in the resolution procedure. We remark that in general we can only
read off $\sum_q h^{p,q} (\pm 1)^{p+q}$ from the \epop; this is sufficient to
determine all Hodge numbers separately only for $n \leq 1$.

\nsection{Old and New Level Rank Dualities} \label{slevr}

We have already seen \erf{ausn} that distinct \lie ic cosets
can describe the same conformal field theory. In fact, this phenomenon occurs
rather frequently: in \cite{fuSc2} the equivalence of four series of \nn\
coset models was established. As these identities arise by exchanging (a
simple function of) the rank and the level of the algebras involved, they are
commonly referred to as `level rank dualities'.

Based on well-known level rank dualities between \wzwts \cite{mnrs}, which are
however {\em not} isomorphisms between these theories, we were able to set up
an isomorphism between the following \nn\ coset models (the notation is taken
from \cite{sche3} and \cite{fuSc}):
\begin{eqnarray}
(B, 2n+1, 2k+1) & \equiv & (B, 2k+1, 2n+1) \\
(B, 2n, 2k+1)   & \equiv & (B, 2k+1, 2n)_{|D} \\
(BB, m+2, 1) & \equiv & (CC, 2, 2m+1)  \\
(CC, n, k) & \equiv & (CC, k+1, n-1)   \label{ccdu}    .
\end{eqnarray}
The subscript $D$ in the last line indicates that one must take the $D$-type
modular invariant for these models rather than the diagonal one.
The last two identities have been suggested in
\cite{kasu2} because of the corresponding symmetry of the central charges.
They have also been observed in the string spectra and in the \epop s in
\cite{sche3} .

In \cite{fuSc2} the identity of these models was established by the
construction of one-to-one correspondence $\Tau$ between all primary fields
(including a careful treatment of the fixed points). As an example, in the
case of the last duality \erf{ccdu} which reads in the full notation
\be \frac{ (C_n)_k \oplus (D_{2n-1})_1 }{(C_{n-1})_{k+1} \oplus \ue_{2(k+n+1)}}
\equiv \frac{ (C_{k+1})_{n-1} \oplus (D_{2k+1})_1 }{(C_k)_n \oplus
\ue_{2(k+n+1)}} \quad ,\ee
$\Tau$ is constructed by relating $(C_n)_k$ and $(C_k)_n$ resp.\
$(C_{n-1})_{k+1}$ and $(C_{k+1})_{n-1}$ by the level rank duality of \wzwts\
\cite{mnrs} and by a prescription for the map on the \ue\ and $D_d$ parts of
the theory.

The mapping $\Tau$ can be shown to preserve both ring structures
present in these models: on the one hand, as $\Tau$ preserves the \smat\ --
and thus the fusion ring -- and respects the conformal dimensions, it is an
intertwiner for the whole modular group. On the other hand, the chiral ring
structure and the superconformal charges are respected, too.
Moreover, reasoning along the lines given in \cite{alts4}
it should not be too hard to show that also the branching functions coincide.

The arguments given in \cite{fuSc2} show that the two respective
theories are exactly identical as conformal field theories, and do not
merely represent different points in the moduli space of one
theory. In fact, changing the moduli
generically also changes the conformal dimensions and even the fusion structure
of the theory, as can be easily seen e.g.\ when looking at the situation at
$c=1$. (An arbitrary marginal deformation of a rational \cft\ does not even
lead to rational \cft.)

We expect that the techniques used in the cases above also allow for an
explanation of the level rank duality in the $A$-series on the level of
Hilbert spaces which was proven in \cite{kasu2} for the symmetry algebras.

\nsection{Conclusions}

One may, of course, extend the analysis presented above and also include
non diagonal modular invariants. For many purposes a complete survey of
the spectra occuring in this class of \nn\ models would be helpful.
One might get a better feeling of whether the old suspicion is true that all
rational \cfts\ are related in some way to coset models, possibly including
additional orbifoldizations.

It is also important to obtain more information about the massless
spectrum of these theories, e.g.\ the number of $E_6$ singlets. One may also
ask whether the models presented in this talk admit a description as
a Calabi-Yau manifold or via a Landau-Ginzburg potential. Finally it would
be interesting to get some insight in whether all coset models with \nn\
superconformal symmetry admit a description in the form \erf{kasuform}.

\vspace{3cm}
{\bf Acknowledgments:} The results presented in this talk are mostly the
outcome of a very pleasant collaboration with J\"urgen Fuchs.
I would also like to thank M.\ Kreuzer, W.\ Lerche, A.N.\ Schellekens
and M.G.\ Schmidt
for stimulating and helpful discussions. Finally I would like to thank the
organizers of the III Conference on Mathematical Physics for their efforts
and for having given me the possibility to present these results.
Financial support from the Studienstiftung des deutschen Volkes is gratefully
acknowledged.

\end{document}